\documentclass[iop]{emulateapj-rtx4}
\usepackage{mathrsfs }
\usepackage{natbib}
\usepackage{multirow}
\usepackage{amssymb}
\usepackage{ulem} 
\newcommand{\NHI}{\ensuremath{N_{\rm HI}}}
\newcommand{\bd}{\ensuremath{{b_{\rm d}}}}
\newcommand{\bdmin}{\ensuremath{{b_{\rm min}}}}
\newcommand{\ValT}{\ensuremath{1.94 \pm 0.05}}
\newcommand{\Valb}{\ensuremath{17.9 \pm 0.2}}
\newcommand{\Valgamma}{\ensuremath{0.46 \pm 0.05}}
\newcommand{\ValGgamma}{\ensuremath{0.15 \pm 0.02}}

\def\kms{km~s$^{-1}$}
\def\cm2{cm$^{-2}$}

\def\ltsima{$\; \buildrel < \over \sim \;$}
\def\gtsima{$\; \buildrel > \over \sim \;$}
\def\simgt{\lower.5ex\hbox{\gtsima}}
\def\simlt{\lower.5ex\hbox{\ltsima}}

\usepackage{float}

\begin{document}

\title{The Temperature-Density Relation in the Intergalactic Medium at redshift $\langle z \rangle =2.4$ }
\author{
 Gwen C. Rudie,\altaffilmark{2}
 Charles C. Steidel,\altaffilmark{2} \&
Max Pettini\altaffilmark{3} 
 }

\altaffiltext{1}{Based on data obtained at the W.M. Keck Observatory, which is operated as a scientific partnership among the California Institute of Technology, the University of California,  and the National Aeronautics and Space Administration, and was made possible by the generous financial support of the W.M. Keck Foundation.}
\altaffiltext{2}{Cahill Center for Astronomy and Astrophysics, California Institute of Technology, MS 249-17, Pasadena, CA 91125, USA}
\altaffiltext{3}{Institute of Astronomy, Madingley Road, Cambridge CB3 0HA, UK}

\email{gwen@astro.caltech.edu}


\shortauthors{Rudie, Steidel, \& Pettini}


\shorttitle{The $T-\rho$ Relation in the IGM at $\langle z \rangle =2.4$}

\begin{abstract}
We present new measurements of the temperature-density ($T-\rho$) relation for neutral hydrogen in the $2.0 < z < 2.8$ intergalactic medium (IGM) using a sample of $\sim$6000 individual \ion{H}{1} absorbers fitted with Voigt profiles constrained in all cases by multiple Lyman series transitions. We find model independent evidence for a positive correlation between the column density of \ion{H}{1} (\NHI) and the minimum observed velocity width of absorbers (\bdmin). With minimal interpretation, this implies that the $T-\rho$ relation in the IGM is not ``inverted'', contrary to many recent studies. Fitting \bdmin\ as a function of \NHI\ results in line width - column density dependence of the form $\bdmin = b_0 \left( \NHI/N_{\rm HI,0} \right)^{\Gamma -1}$ with a minimum line width at mean density ($\rho/\bar\rho = 1, N_{\rm HI, 0} = 10^{13.6}$ cm$^{-2}$) of $b_0= \Valb$ km s$^{-1}$ and a power-law index of $(\Gamma -1) = \ValGgamma $. Using analytic arguments, these measurements imply an ``equation of state'' for the IGM at $\langle z \rangle = 2.4$ of the form $T=T_0 \left(\rho/\bar\rho\right)^{\gamma-1}$ with a temperature at mean density  of $T_0= [\ValT] \times 10^4 $ K and a power-law index $(\gamma -1) = \Valgamma$.

\end{abstract}


\section{Introduction}

The ``equation of state'' of the low-density intergalactic medium is believed to be controlled by two principle processes: photo-heating and adiabatic cooling. The cooling is most directly tied to the expansion of the Universe, while the heating is expected to be a complicated mixture of relic effects from the reionization of hydrogen and helium plus the current heating, predominantly from the UV background \citep{hui97,sch99}. This naturally imposes a relationship between the temperature and density of intergalactic gas. Denser gas is expected to trace larger over-densities, for which adiabatic cooling is suppressed because such regions are more bound against the expansion of the Universe. At the same time, denser gas has a much faster recombination time scale and thus presents a larger cross-section for photo-ionization which leads denser gas to experience greater heating.

Recently, this simple picture has been called into question on the basis of several statistical studies of the transmitted flux observed in individual pixels within the Lyman-$\alpha$ forest \citep{bec07,bol08,vie09,lid10,cal12,gar12}. These studies require comparison with large numerical simulations in order to interpret the physical implications of the detailed shape of these flux distributions. Further, the shape of the distribution may be sensitive to the full thermal history of the gas, not just the present temperature-density ($T-\rho$) relation \citep{pee10}. 

\citet{bec07} considered the transmitted flux probability distribution function (PDF) and compared it to the numerical models of \citet{mir00}, finding that none of the models could explain the observations. They found that a $T-\rho$ relation in which lower density regions were \textit{hotter} (an inverted $T-\rho$ relation) provided a better fit to the data. Since this study, several other authors have obtained similar results using transmitted flux PDF analysis \citep{bol08,vie09,cal12}, wavelet analysis \citep{lid10}, or a combination of both \citep{gar12}.

In contrast, early work on the $T-\rho$ relation and its evolution with redshift was performed via Voigt profile fitting of individual \ion{H}{1} absorption lines. These early studies found a monotonic $T-\rho$ relation with a positive power-law index \citep{sch00,ric00,bry00,mcd01}. 

In this letter, we return to the more direct test of the $T-\rho$  relation in the IGM using Voigt profile fits to individual \ion{H}{1} absorbers. This method relies on the expected relationship between the column density of neutral hydrogen of an absorber, \NHI, and its local overdensity, $\rho/\bar{\rho}$ \citep{dav99, sch01}. If such a relationship exists, then one expects to observe a correlation between \NHI\ and the velocity widths of individual thermally broadened absorbers, \bd. Absorbers having only thermal broadening\footnote{Other sources of broadening of absorbers include bulk motions of the gas (generally parametrized as a turbulent component of \bd) and differential Hubble flow which broadens the absorption lines originating from the most diffuse and physically extended absorbers.} are expected to have the smallest \bd\ at a given \NHI, \bdmin. Thus, by observing the behavior of the low-\bd\ ``edge'' of the distribution of \bd\ versus \NHI, it is possible to infer the relationship between $T$ and \NHI\ and through theory, the IGM $T-\rho$ relationship. 

Here we present analysis of individual absorbers fitted in 15 high-resolution, high-S/N spectra of luminous QSOs at $2.5 < z < 2.9$ from the Keck Baryonic Structure Survey (KBSS; \citealt{gcr12}). We discuss the data set used in this study as well as the line-fitting procedure in Section \ref{data}. In Section \ref{b-NHI}, the fit to the minimum \bd\ (\bdmin) as a function of \NHI\ is presented, followed by Section \ref{results} in which the physical implications of the results in the context of the $T-\rho$  relation are discussed. The results and their implications are summarized in Section \ref{con}.

Throughout this paper we assume a $\Lambda$-CDM cosmology with $H_{0} = 70$ \kms\  Mpc$^{-1}$, $\Omega_{\rm m} = 0.3$, and $\Omega_{\Lambda} = 0.7$.

\section{Data and Analysis}
\label{data}

The data analyzed in this paper are taken from the Keck Baryonic Structure Survey (KBSS). The KBSS is a large spectroscopic survey designed to study the gaseous surroundings of high-$z$ star-forming galaxies by combining redshift surveys for galaxies  at $2 < z < 3$ in the fields of bright QSOs with absorption line data from those same QSOs. The data include high-resolution, high signal-to-noise echelle spectra of 15 luminous ($m_V \simeq 15.5-17$) QSOs located near the centers of the KBSS fields. The KBSS will be described fully by Steidel et al. (in prep); however, the data of relevance to this paper as well as the analysis of the absorption-line sample have been presented in \citet{gcr12}.  Here we give a brief summary. 

 The QSO spectra were obtained with the High Resolution Echelle Spectrometer \citep[HIRES;][]{vog94} on the Keck I telescope.  The HIRES spectra have $R\simeq 45,000$ (FWHM$\simeq 7$ \kms) and S/N $\sim 50-200$ per pixel. They cover the wavelength range 3100 -- 6000 \AA\ with no spectral gaps, allowing for the observation of Ly$\beta$ $\lambda 1025.7$ down to at least $z= 2.2$ in all 15 of our sightlines. The additional constraints provided by Ly$\beta$ (and in many cases, additional Lyman series transitions) allow for highly accurate measurements of column densities and (most importantly for this paper) line widths for the \ion{H}{1} absorbers in our sample. 

\begin{deluxetable}{llcrc}
\tablecaption{KBSS Absorption Line Sample}  
\tablewidth{0pt}
\tablehead{
\colhead{Name} & \colhead{$z_{\rm QSO}$\tablenotemark{a}} &  \colhead{$z_{\rm Ly\alpha}$ range} &  \colhead{S/N\tablenotemark{b}} & \colhead{S/N\tablenotemark{b}}\\
\colhead{} & \colhead{} &\colhead{} &\colhead{Ly$\alpha$} &\colhead{Ly$\beta$} }
\startdata
 Q0100+130 (PHL957) &  2.721 &   2.0617-- 2.6838 &   ~77   &   50  \\
 HS0105+1619 & 2.652 &   2.1561-- 2.6153 &  127  	&   89  \\
 Q0142$-$09 (UM673a) &  2.743 &  2.0260-- 2.7060   & ~71  	&   45  \\
 Q0207$-$003 (UM402) &  2.872 & 2.1532-- 2.8339  &   ~82  	&   55  \\
 Q0449$-$1645 & 2.684 &   2.0792-- 2.6470 &  ~73  	&   41  \\
 Q0821+3107 &   2.616 &   2.1650-- 2.5794  &    ~50  	&   33  \\
 Q1009+29 (CSO 38) &   2.652 &    2.1132-- 2.6031& ~99  	&   58  \\
 SBS1217+499 &   2.704  &    	 2.0273-- 2.6669     &  ~68  	&   38  \\
 HS1442+2931 &   2.660 &  2.0798-- 2.6237   &  ~99  	&   47  \\
 HS1549+1919 &   2.843 &    2.0926-- 2.8048  &  173  	&   74  \\
HS1603+3820 &  2.551 & 2.1087-- 2.5066  &  108  	&   58  \\
 Q1623+268 (KP77) &   2.535 &     2.0544-- 2.4999  &    ~48  	&   28  \\
 HS1700+64 &  2.751 &    2.0668-- 2.7138  &      ~98  	&   42  \\
 Q2206$-$199 &  2.573 &  2.0133-- 2.5373   &  ~88  	&   46  \\
 Q2343+125 & 2.573 &   2.0884-- 2.5373 &                     ~71  	&   45      
 \enddata
 \tablenotetext{a}{The redshift of the QSO \citep{tra12}}
 \tablenotetext{b}{The average signal to noise ratio per pixel of the QSO spectrum in the wavelength range pertaining to Ly$\alpha$ and Ly$\beta$ absorption.}
     \label{field}
\end{deluxetable}

The reduction of the QSO spectra and a detailed description of the process of fitting the forest are discussed in \citet{gcr12}. The final Voigt profiles fits to the full Ly$\alpha$ and Ly$\beta$ forests of the 15 KBSS QSOs were completed using VPFIT\footnote{http://www.ast.cam.ac.uk/$\sim$rfc/vpfit.html; \copyright ~2007 R.F. Carswell, J.K. Webb} written by  R.F. Carswell and J.K. Webb. The redshift range included in the fit is given in Table \ref{field}. In this paper, we use the pathlength weighted mean redshift of the sample $\langle z \rangle = 2.37$ as the fiducial redshift \citep{bah69}.\footnote{The comoving pathlength, $dX$, was defined by \citet{bah69} such that absorbers with constant physical size and comoving number density would have a contant number per $dX$.}

The final \ion{H}{1} absorber catalog includes 5758 absorbers with $12.0 < \log(\NHI/ \textrm{cm}^{-2}) < 17.2$ and $ 2.02 < z <  2.84$ over a total redshift path length of $\Delta z = 8.27$. 
This \ion{H}{1} sample is the largest ever compiled at these redshifts and is more than an order of magnitude larger than previous samples that included constraints from higher-order Lyman lines. 

\section{The Temperature-Density Relation in the IGM}

\label{b-NHI}

The equation of state of the IGM is expected to have the form
\begin{equation}
T=T_0 \left( \frac{\rho}{\bar{\rho}} \right)^{\gamma-1}
\end{equation}
where $T_0$ is the temperature at the mean mass density ($\overline \rho$) \citep{hui97}. 

Under the assumption of a relatively uniform ionizing radiation field, a power-law relationship between \NHI\ and $\rho$ is also expected. \citet{sch01} presented a model for the low-density IGM in which `clouds' are in local hydrostatic equilibrium and therefore typically have sizes comparable to the local Jeans length. Employing this assumption, \citet{sch01} derived a relationship between \NHI\ and local overdensity, $\rho/\overline \rho$. Using updated cosmology and evaluating at the path-length weighted mean redshift of the sample: 
\begin{equation}
\label{jeans}
\rho_b/\overline \rho_b  \approx \left(\frac{N_{\textrm{\footnotesize{HI}}}}  {10^{13.6}}\right)^{2/3}T_4^{0.17} \left( \frac{\Gamma_{12}}{0.5}   \right)^{2/3} \left( \frac{1+z}{3.4} \right)^{-3},
\label{schaye}
\end{equation}
where $T_4$ is the gas temperature in units of $10^4$ K and $\Gamma_{12}$ is the hydrogen photoionization rate in units of $10^{-12}$ s$^{-1}$ with the normalization suggested by \citet{fau08}. Assuming this scaling\footnote{We note here that with $T_4 = 2$ as we find in this paper, the \NHI\ corresponding to $\bar{\rho}$ is $\log(\NHI/$\cm2$)=13.7$, a change of 0.1 dex. However, the uncertainty in other parameters (e.g. $\Gamma_{12}$) is large enough, that we do not consider this small effect in this paper.}, absorbers with $\log(\NHI/ \textrm{cm}^{-2}) \approx 13.6$ are expected to trace gas at the mean density of the universe at $z=2.4$. 

\begin{figure}
\center
\includegraphics[width=0.45\textwidth]{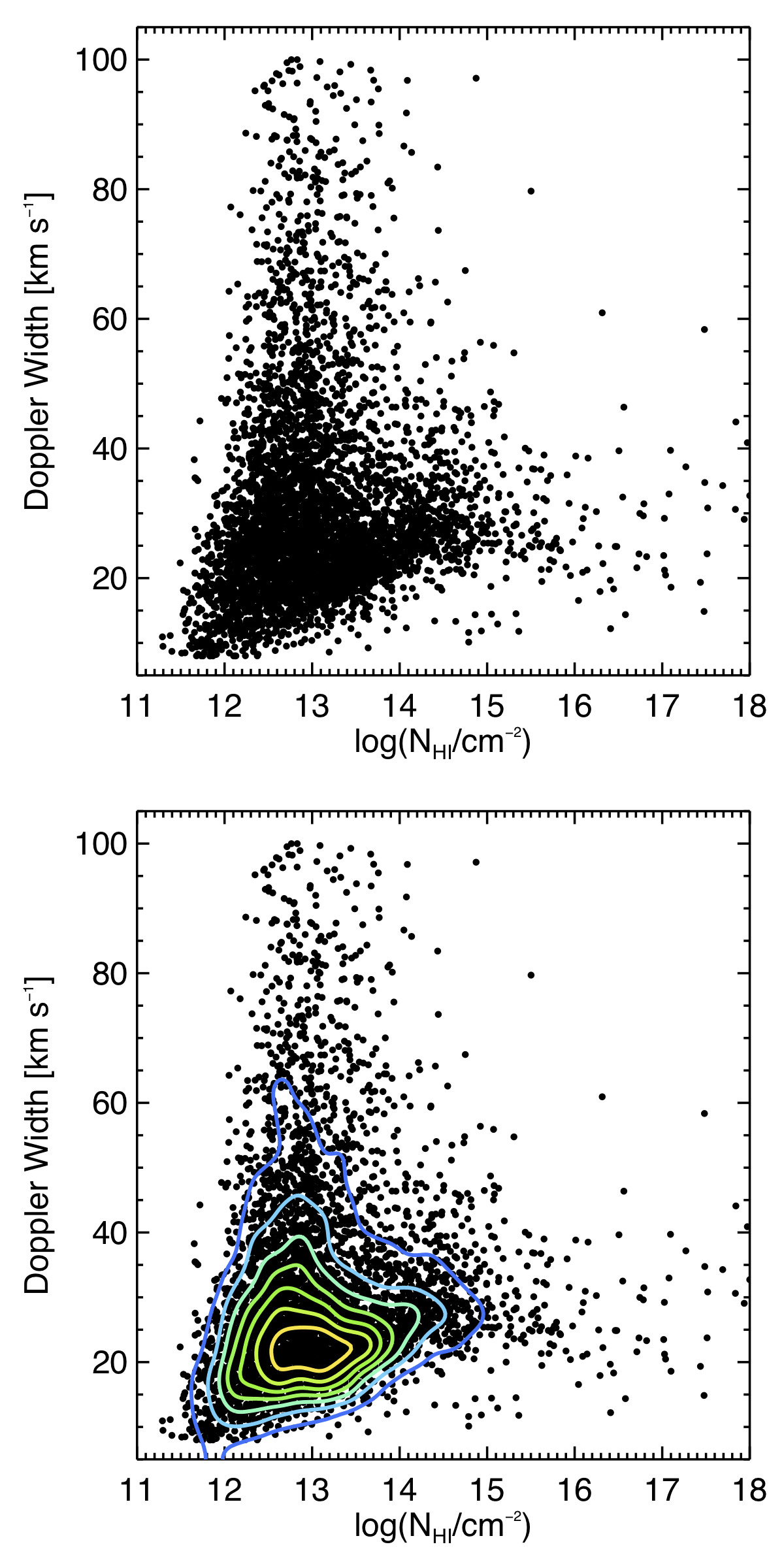}
\caption{\small The doppler widths of absorbers (\bd) versus their column density (\NHI) for all the absorbers in the HI sample with relative errors in \NHI\ and \bd\ less than 50\%. Point-density contours are over plotted to guide the eye in the bottom panel. Note that the minimum value of \bd\ at each \NHI\ increases with increasing \NHI\ suggesting a normal $T-\rho$ relation.}
\label{b_vs_N}
\end{figure}

Thermally broadened absorbers are also expected to follow a power law relation between \bd\ and temperature: $\bd~\propto~T^{1/2}$. Combining with the expected $T-\rho$ relation and the conversion between \NHI\ and $\rho$, the relationship between \bdmin\ and \NHI\ would be a power law of the form:
\begin{equation}
\bdmin = b_0 \left(\frac{\NHI}{N_{\rm{HI, 0}}}\right)^{\Gamma -1}
\label{eqn_b_n}
\end{equation}
where $b_0$ is the minimum line width of absorbers with $\NHI = N_{\rm HI,0}$.  With this formalism, $(\gamma -1 )$ is proportional to $(\Gamma -1)$ \citep[see eg.][]{sch99}.

More explicitly, for pure thermal broadening:
\begin{equation}
b= (2 k_B T/m_{\rm p})^{1/2}
\label{bT}
\end{equation}
where $k_B$ is the Boltzmann constant and $m_{\rm p}$ is the mass of the proton. This suggests:
\begin{equation}
\log  \left( T \right)= C + 2 \log\left( \frac{b}{\textrm{km s}^{-1}} \right)
\label{eqn_Tb}
\end{equation}
where
\begin{equation}
C=\log \left(  \frac{m_{\rm p}}{2 k_B} \frac{(\textrm{km s}^{-1})^2}{\textrm{K}}\right) = 1.78.
\end{equation}

Rearranging the above equations, we expect the conversion between the index of $\bdmin(\NHI)$ and the $T - \rho$ relationship to be roughly:\footnote{In the following equation, we have neglected the minor dependence of $\rho_b/\overline \rho_b$ on $T$ from equation \ref{schaye}.}
\begin{equation}
\gamma-1\approx 3 \left( \Gamma -1\right)
\label{eqn_gamma}
\end{equation}

In the sections that follow, we fit to the trend of the \bdmin\ as a function of \NHI\ for the data sample. We then use equations \ref{eqn_Tb} and \ref{eqn_gamma} to estimate the $T-\rho$ relation in the $\langle z \rangle =2.4$ IGM.

\begin{figure*}
\centerline{
\includegraphics[width=0.4\textwidth]{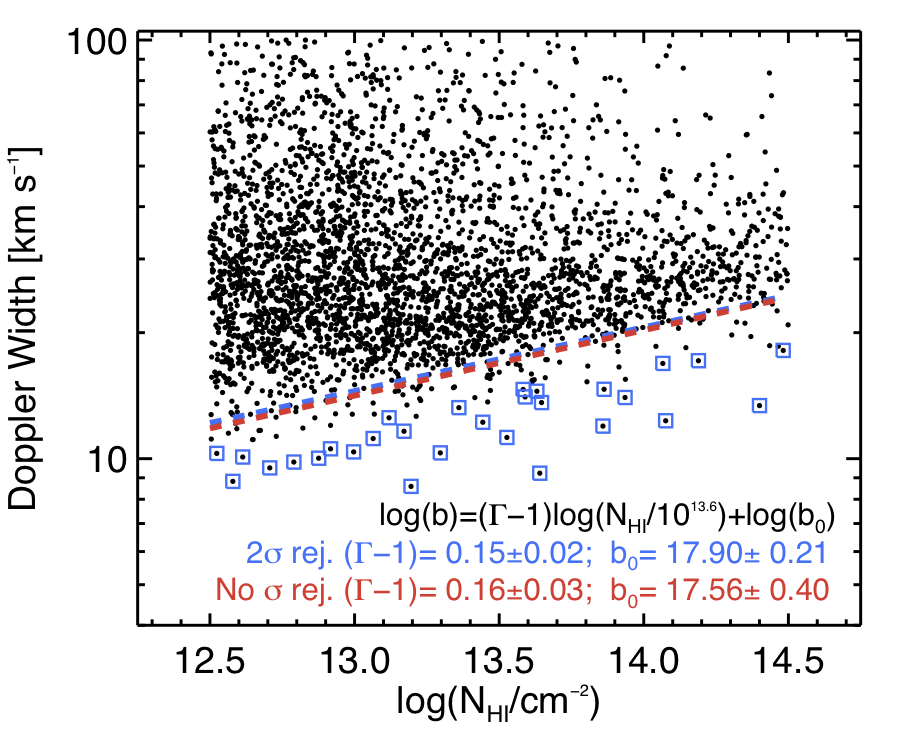}\includegraphics[width=0.65\textwidth]{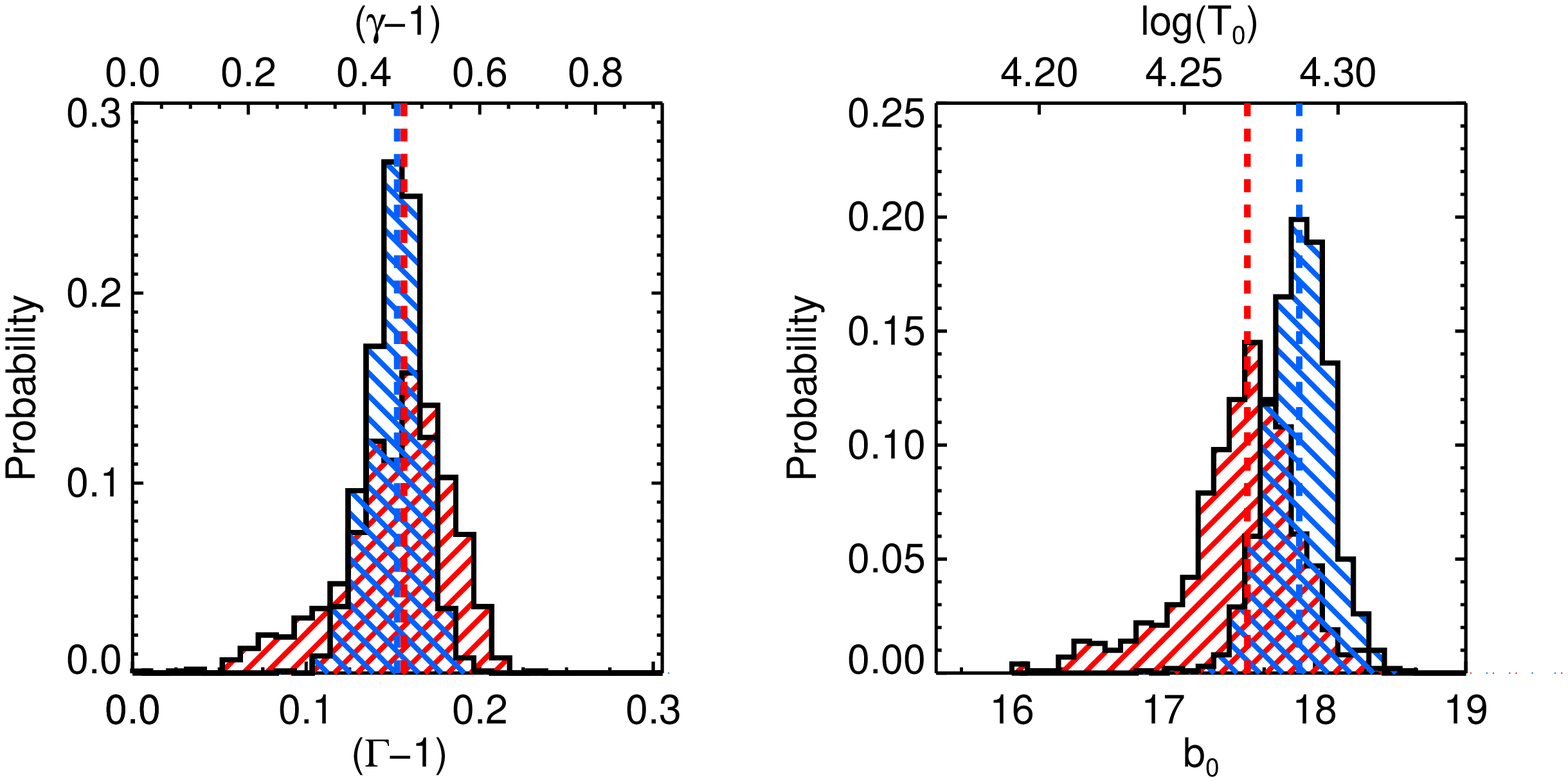}}
\centerline{
\includegraphics[width=0.4\textwidth]{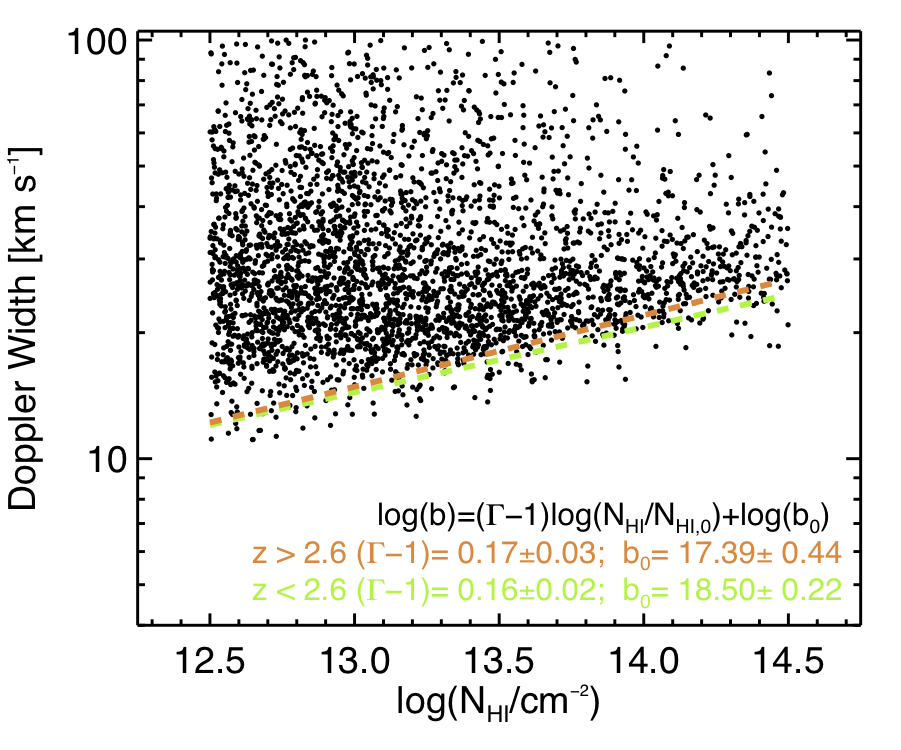}\includegraphics[width=0.65\textwidth]{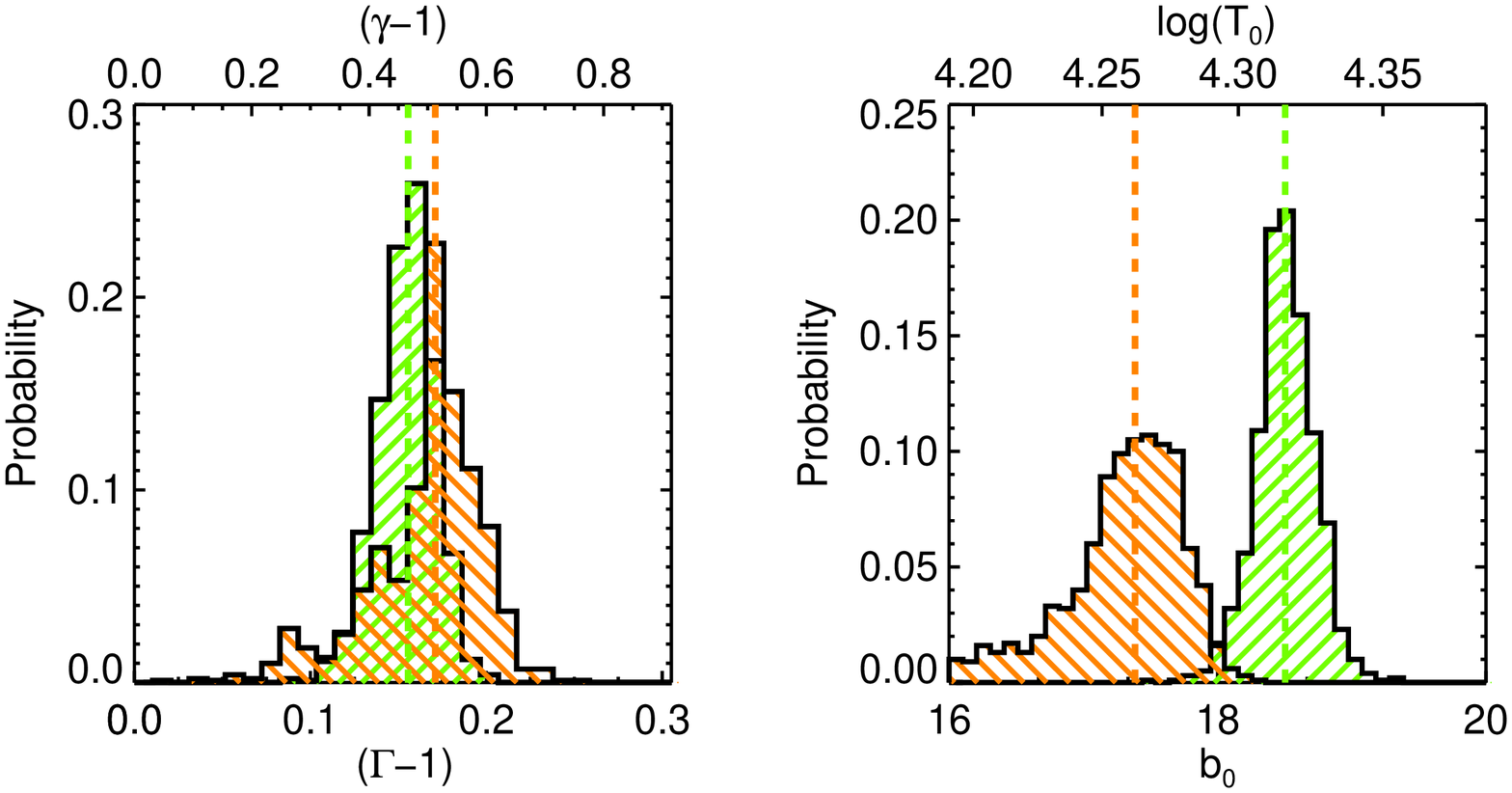}}
\caption{\small  \textit{Left panel:} \bd\ versus \NHI\ for all absorbers in the KBSS HI sample with $12.5 < \log(\NHI/ \textrm{cm}^{-2}) < 14.5$ and relative errors in \bd\ and \NHI\ less than 50\%(black points).  The curves show the resulting best power-law fits determined by applying the bootstrap resampling method to the sample followed by the iterative power-law fit method described by \citet{sch99}. \textit{Top panels:} The blue and red lines pertain to the data set with the $\sigma$-rejected absorbers (blue boxes) excluded or included respectively. \textit{Top Center and Right:} The probability distribution for the fitted parameters using the 50\% relative-error rejection algorithm (red) and both the 50\% relative-error rejection algorithm and the $2\sigma$ rejection algorithm (blue). \textit{Bottom panels:} The orange and green curves pertain to the $z>2.6$ and $z<2.6$ subsamples of the $\sigma$-rejection data set. Here $N_{\rm HI,0}$ varies between the subsamples with  $\log(N_{\rm HI,0}/$\cm2$)=13.7$ for $z < 2.6$ and $\log(N_{\rm HI,0}/$\cm2$)=13.4$ for $z > 2.6$ due to the redshift dependence in equation \ref{schaye}. Note that the \bdmin-\NHI\ relation is roughly constant, but the expected $N_{\rm HI,0}$ evolution would require evolution in $b_0$ and $T_0$.}
\label{bN}
\end{figure*}

\subsection{A ``normal'' $T-\rho$ relation}

The measured values of \bd\ and \NHI\ considered in this letter are presented in Figure \ref{b_vs_N}. Examination of Figure \ref{b_vs_N} already suggests the main result of this paper; focusing on the low-\bd\ edge of the distribution, one can easily see that the lower envelope increases monotonically with \NHI. Such behavior is expected for a \textit{normal} temperature-density relationship in which denser regions are hotter.  

\subsection{Outlier Rejection}

\label{outlier}

Before fitting to the ``ridge-line'' of \bdmin\ versus \NHI, we exclude those absorbers that have large errors in their measured parameters. \citet{sch99} suggest a simple algorithm for the outlier rejection in which absorbers with relative error in \bd\ or \NHI\  larger than 50\% are excluded.\footnote{The relative error considered is computed from the formal error bars output by VPFIT for both \bd\ and \NHI.} This method is expected to primarily reject lines which originate in blends and are thus unlikely to have accurate line-width measurements. Additionally, we exclude those absorbers with \bd = 8 \kms and \bd = 100 \kms. These line widths correspond to the allowed line-width limits input to VPFIT and as such are artificial. The absorbers which remain after those exclusions are shown in Figure \ref{b_vs_N}.


Considering Figure \ref{b_vs_N}, we note a small set of points at very low \bd\ which seem to lie significantly below the main locus of points. These absorbers are likely to be unidentified metal line contaminants, and thus we consider a $\sigma$-rejection algorithm to exclude them. 

The $\sigma$-rejection algorithm is applied as follows. Only absorbers with $\bd < 40$ \kms\ are considered. The data are sorted by their \NHI.  The absorbers are placed into \NHI\ bins of width 0.25 dex and the mean and standard deviation of \bd\ are computed for the bin. Absorbers with \bd\  2$\sigma$ or larger from the mean are flagged and excluded. This process is iterated; the set of absorbers excluded via this method with \bd\ lower than the mean are surrounded by (blue) boxes in Figure~\ref{bN} and are excluded from the fitting procedure. We have checked the majority of $\sigma$-rejected absorbers by hand and find that they are either previously unidentified metal lines or parts of blended saturated \ion{H}{1} systems whose individual properties are poorly constrained.

In the following section, we proceed with the measurement of the \bdmin -- \NHI\  relation. We show that the application of the $\sigma$-rejection algorithm does not produce significant changes to the results. In the conclusions of the paper, however, we prefer the combined error and $\sigma$-rejection method as it excludes the majority of the poorly measured absorbers as well as those which lie far below the \bdmin\ -- \NHI\ trend. 

\subsection{Fitting \bdmin\ versus \NHI}

\label{Schaye_power_law}

In this section we measure the \bdmin\ -- \NHI\ relation using the two data sets resulting from the outlier rejection methods described above. To fit the trend between \bdmin\  and \NHI, we follow the iterative power-law method proposed by \citet{sch99}. A power-law relationship of the form shown in equation \ref{eqn_b_n} with $\log(N_{\rm HI,0}/$\cm2)$ = 13.6$ is fit to the data. Data points which have \bd\ larger than one mean absolute deviation above the fit are discarded. The power law is refit to the remaining absorbers and the rejection and refitting are repeated until the power law converges. Then, absorbers more than one mean absolute deviation below the fit are removed \textit{once}, and the power law is refit. This fit is taken as the final form of the minimum \bdmin\ -- \NHI\ relation. Errors in the parameters of the fit are derived by applying the bootstrap resampling method to the data sample and following the above outlined procedure.

\begin{deluxetable*}{lccccccccc}
\scriptsize
\tablecaption{Fits to the $\bdmin - \NHI$ and $T-\rho$ relation in IGM}  
\tablewidth{0pt}
\tabletypesize{\scriptsize}
\tablehead{
\colhead{Outlier Rej.} &\colhead{$z$ range} & \colhead{$\langle z \rangle$} & \colhead{$\log(N_{\rm HI,0}$/\cm2)} & \colhead{$(\Gamma - 1)$} &  \colhead{$(\gamma - 1)$\tablenotemark{a}}  &\colhead{$b_0$}  & \colhead{$T_0/10^4$ K\tablenotemark{b}}}
\startdata
default & $2.0-2.8$ & 2.4 &13.6 & 0.156 $\pm$ 0.032 &  0.47 $\pm$ 0.10 &  17.56 $\pm$ 0.40 & 1.87 $\pm$ 0.08 \\
$2\sigma$-rej. & $2.0-2.8$ & 2.4 & 13.6 & 0.152 $\pm$ 0.015 &  0.46 $\pm$ 0.05 &  17.90 $\pm$ 0.21 & 1.94 $\pm$ 0.05\\\\
\hline \\
$2\sigma$-rej. & $2.0 - 2.6$ & 2.3 & 13.7 & 0.156 $\pm$ 0.016 &  0.47 $\pm$ 0.05 &  18.50 $\pm$ 0.22 & 2.07 $\pm$ 0.05 \\
$2\sigma$-rej. & $2.6 - 2.8$ & 2.7 & 13.4 & 0.171 $\pm$ 0.032 &  0.51 $\pm$ 0.10 &  17.39 $\pm$ 0.44 & 1.83 $\pm$ 0.09
 \enddata
     \label{igm_temp}
     \tablenotetext{a}{$(\gamma - 1)$ is calculated from the measured value of $(\Gamma - 1)$ using equation \ref{eqn_gamma}.}
    \tablenotetext{b}{$T_0$ is calculated from the measured value of $b_0$ using equation \ref{eqn_Tb}.}
   \end{deluxetable*}

\begin{deluxetable*}{llcccl}
\scriptsize
\tablecaption{Comparison with Previous measurements of the $T-\rho$ relationship}  
\tablewidth{0pt}
\tabletypesize{\scriptsize}
\tablehead{
\colhead{Source} &  \colhead{Method} & \colhead{$\langle z \rangle$}  & \colhead{$T_0/10^{4} \rm K$} & \colhead{$(\gamma -1)$}  & \colhead{Comment}   }
\startdata
\multirow{2}{*}{\citet{sch00}} & Line Fitting & 2.3    &  $1.17$ &    	0.27  & \multirow{2}{*}{Their Figure 6}\\
& Line Fitting & 2.5	&  $1.38$     &	0.56   & \\\\
\multirow{2}{*}{\citet{ric00}}  & Line Fitting & 2.75 &  $2.52$ & $0.22 \pm 0.10$  & \multirow{2}{*}{Their Table 6}\\
& Line Fitting & 1.90 &  $ 1.77 $ & $0.32 \pm 0.30$ & \\\\
 \citet{bry00}\tablenotemark{b} & Line fitting & 2.7 & $1.65 $\tablenotemark{a}& $0.29$ & \\
 \citet{mcd01} & Line Fitting & 2.41 &  $1.74 \pm 0.19 $ & $ 0.52 \pm 0.14$ & Their Table 5, $z_{\rm sim}=3$\\
 \citet{zal01} & Transmission Power Spectra & 2.4 & $1-3 $ &  0 to 0.6 &  Their Figure 3\\
 \citet{bec07} &  Flux PDF & $2-6$ & \nodata & $\approx -0.5$ &  \\\\
  \multirow{3}{*}{\citet{bol08}\tablenotemark{c} }& Flux PDF & 2.07& \nodata  & $-0.33$ & \\
  & Flux PDF & 2.52 & \nodata  &  $-0.46$ & \\
& Flux PDF &  2.94 & \nodata  & $-0.56$ & \\\\ 
 \citet{vie09} &  Flux PDF & 3 &  $1.9 \pm 0.6$ & $-0.25 \pm 0.21$ & \\\\
   \multirow{2}{*}{\citet{lid10}} & Wavelet Analysis &   2.6 & $1.6 \pm .6 $\tablenotemark{d} & \nodata  &  \multirow{2}{*}{Their Figure 30}\\
& Wavelet Analysis &   2.2 & $2.1 \pm .7 $\tablenotemark{d} & \nodata &\\\\
\citet{bec11} & Curvature Analysis &  $2.4$ & $1.11 \pm 0.06$\tablenotemark{e}& \nodata &Their Table 3\\\\
\multirow{6}{*}{\citet{gar12}\tablenotemark{f}} & Wavelet Analysis + Flux PDF & 2.1 & $1.7 \pm 0.2 $ &$0.11 \pm 0.11$ &\multirow{6}{*}{Their Table 2} \\
& Wavelet Analysis  & 2.1 & $1.6 \pm 0.5 $ &$> -0.14$  & \\
& Flux PDF & 2.1 & $1.5 \pm 0.3 $ &$-0.01 \pm 0.14$  & \\
 & Wavelet Analysis + Flux PDF & 2.5 & $1.3 \pm 0.4 $ & $ >-0.05$ & \\
& Wavelet Analysis  & 2.5 & $1.6 \pm 0.4$& $ >-0.08$ & \\
& Flux PDF & 2.5 & $1.4 \pm 0.9 $ & $>-0.31$ & \\\\
\multirow{2}{*}{\citet{cal12}} & Flux PDF & 3 & $1.93 \pm 0.48$ & $-0.10\pm0.21$ & \multirow{2}{*}{Their Table 5} \\
& Flux PDF\tablenotemark{g} &  $3$ & $1.79 \pm 0.35$  & $-0.30\pm 0.12$ &  \\\\
\hline\\
\multirow{3}{*}{This Work} & \multirow{3}{*}{ Line fitting} &   \multirow{3}{*}{2.37} & \multirow{3}{*}{\ValT}  & \multirow{3}{*}{\Valgamma}  & Results from data set \\
& & & & &using both relative error \\
& & & & &and $\sigma$-rejection
 \enddata
     \label{prev_res}
     \tablenotetext{a}{The authors suggest this is an upper limit.} 
      \tablenotetext{b}{Based on the spectrum of HS 1946$+$7658}
       \tablenotetext{c}{Measured by \citet{kim07}, sample includes 18 QSOs (metals removed by hand to eliminate noise in flux PDF.)}
     \tablenotetext{d}{Marginalized over $(\gamma -1 ) =0-0.6$}
     \tablenotetext{e}{Assuming $\gamma -1 = 0.56$}
    \tablenotetext{f}{Find that flux PDF results in lower values of $\gamma$ than wavelet analysis in general. }
     \tablenotetext{g}{Joint analysis with \citet{kim07} .}
\end{deluxetable*}

Figure \ref{bN} shows the results of the iterative power-law fit for both data sets formed via the outlier rejection algorithms. The red curves correspond to the fit including the $\sigma$-rejected points, while the blue curves correspond to those without. The center and right panels show the distribution of fit parameters obtained through bootstrap resampling.  As expected, those fits which exclude the low-\bd\ outliers via $\sigma$-rejection (blue histograms) result in tighter distributions for the fitted parameters; however, the best fit values are nearly indistinguishable. 

The best fit parameters $b_0$ and $(\Gamma - 1)$ and their relative errors, as well as their physical counterparts $T_0$ and $(\gamma-1)$, are presented in Table \ref{igm_temp}. $T_0$ and $(\gamma-1)$ are calculated assuming equations \ref{eqn_Tb} and \ref{eqn_gamma} respectively. Notably, both outlier rejection algorithms produce statistically consistent results suggesting that the measurements are robust to changes in the rejection algorithm; however, (as expected) slightly higher $b_0$ are preferred by the algorithms with $\sigma$-rejection. 

Henceforth, we consider the results from the data set formed with both the error rejection and $\sigma$-rejection algorithms applied prior to the power-law fit to \bdmin(\NHI).

\section{Results}

\label{results}

As discussed above, the mapping of \bdmin(\NHI) to the equation of state for intergalactic gas requires a theoretical framework. \citet{sch00} considered simulations of the IGM and found that the minimum values of \bd\ at each \NHI\ were larger than expected given purely thermal broadening. They suggested that all absorbers in the \NHI\ range used to measure the $T-\rho$ relationship experienced mild Hubble broadening (also expected to scale as $T^{1/2}$ for clouds having sizes similar to the local Jeans scale.) In this case, estimates made using the above equations would mildly over-predict the temperature at mean density. We prefer to report physical values tied closely to our measurements rather than rely directly on a specific set of simulations; however, our measurements of $b_0$ and $\Gamma-1$ can be converted to physical parameters using a different set of assumptions. Our simplifying assumption of pure thermal line broadening for absorbers near \bdmin(\NHI) results in $T_0$ values slightly higher than some previous studies which used simulations as a reference (see Table \ref{prev_res}).

In Table \ref{prev_res}, we compare our results with those of previous studies. We find generally good agreement between our results and those of past studies which used line fitting. We further emphasize that our results differ significantly from those obtained using the transmitted flux PDF and similar statistical methods, calling into question the interpretation of such techniques. 

\subsection{\ion{He}{2} Reionization}

One of the expected results of \ion{He}{2} reionization, believed to occur at $z\approx3$, is a significant change in the $T-\rho$ relation due to excess photoionization heating \citep{hui97}. These changes are not long lived, and the exact effects depend on the speed and patchiness with which reionization proceeds and the spectral hardness of the ionizing sources (see \citealt{mcq09}). While detailed discussion is beyond the scope of this letter, we briefly mention that the absorbers in the KBSS \ion{H}{1} sample show no strong evidence for a change in the slope of $\bdmin(\NHI)$ for higher-redshift absorbers, which (presumably) are temporally closer to the \ion{He}{2} reionization epoch. The bottom panels of Figure \ref{bN} show the iterative power law fit to absorbers with, respectively, $z < 2.6$ (green) and $z > 2.6$ (orange) from the sample with $\sigma$-rejection. Notably, the $\bdmin-\NHI$ relation appears to be independent of redshift (left panel); however, the expected evolution in the value of \NHI\ at the mean density (equation \ref{schaye}) results in differing values of  $N_{\rm HI,0}$ for the two redshift bins. This in turn leads to expected variation in the values of $b_0$ and $T_0$ with $z$. The implied parameters from the fit are listed in Table \ref{igm_temp}.

\section{Conclusions}

\label{con}

We have inferred the $T-\rho$ relationship in the $\langle z \rangle = 2.4$ IGM using Voigt profile fitting of individual Lyman lines. Fitting the trend of the minimum line width (\bdmin) as a function of column density (\NHI) with a power law, we find best fit values of \bdmin\ at mean density (which at $\langle z \rangle = 2.4$
corresponds to $N_{\rm HI, 0} = 10^{13.6}$ \cm2) $b_0=\Valb$ km s$^{-1}$ and a power-law index $(\Gamma -1) = \ValGgamma $. Assuming a monotonic relation between \NHI\ and $\rho/\bar{\rho}$, these data support the conclusion that the temperature-density ($T-\rho$) relationship in the IGM is not inverted at $\langle z \rangle = 2.4$ but instead has a power-law index $(\gamma -1) = \Valgamma$. Further, our results suggest a temperature at mean density of $T_0=[\ValT] \times 10^4 $K. Within our sample spanning the redshift range $2.0 \lesssim z \lesssim2.8$, there is no evidence for significant evolution in the $\bdmin - \NHI$ relation or in the power-law index ($\gamma - 1$) of the $T-\rho$ relation.

\acknowledgements
We thank George Becker for his early interest in the $T-\rho$ relation results from the KBSS sample, Joop Schaye who provided helpful advice and insightful comments, and Allison Strom and Ryan Trainor for their careful reading of early drafts. Thanks also to Olivera Rakic for her contributions to the reduction of the QSO data set and for her pertinent advice. The authors wish to acknowledge Ryan Cooke who contributed the fits to the damped profiles in our QSO spectra. 

We wish to acknowledge the staff of the the W.M. Keck Observatory whose efforts insure the telescopes and instruments perform reliably. We thank those of Hawaiian ancestry on whose sacred mountain we are privileged to be guests. 

This work has been supported by the US National Science Foundation through grants AST-0606912 and AST- 0908805. CCS acknowledges additional support from the John D. and Catherine T. MacArthur Foundation and the Peter and Patricia Gruber Foundation. This research has made use of the Keck Observatory Archive (KOA). 

\textit{Facilities:} \facility{Keck:I (HIRES)}

\bibliographystyle{apj}
\bibliography{ms}

\end{document}